\documentclass[aps,pre,floatfix,showpacs]{revtex4}
\usepackage{graphicx}
\usepackage{amsmath}
\begin{document}

\title{Dynamics-Driven Evolution to Structural Heterogeneity in Complex Networks}

\author{Zhen Shao and Haijun Zhou}

\affiliation{Institute of Theoretical Physics, Chinese Academy of Sciences, {P.~O.~Box} 2735, Beijing 100190, China}

\date{\today}

\begin{abstract}
The mutual influence of dynamics and structure is a central issue in complex systems. In this paper we study by simulation slow evolution of network under the feedback of a local-majority-rule opinion process. If performance-enhancing local mutations have higher chances of getting integrated into its structure, the system can evolve into a highly heterogeneous small-world with a global hub (whose  connectivity is proportional to the network size), strong local connection correlations and power law-like degree distribution. Networks with better dynamical performance are achieved if structural evolution occurs much slower than the network dynamics. Structural heterogeneity of many biological and social dynamical systems may also be driven by various dynamics-structure coupling mechanisms.
\end{abstract}

\maketitle

\section{Introduction}

The underlying networks of many biological and social complex systems are distinguished from purely
random graphs. These real-world networks often have the small-world property
\cite{Watts-Strogatz-1998} and scale-free (power-law) vertex degree profiles
\cite{Barabasi-Albert-1999};
they have system-specific local
structural motifs \cite{Milo-etal-2002} and often are organized into communities
\cite{Ravasz-etal-2002,Girvan-Newman-2002}
of different connection densities. In recent years models have been proposed to understand
the structural properties of real-world complex systems
\cite{Albert-Barabasi-2002,Dorogovtsev-Mendes-2002};
among them the $``$rich-get-richer" mechanism of network growth by preferential attachment
\cite{Barabasi-Albert-1999}
gained great popularity. As the connection pattern affects considerably
functions of a networked system, there may exist various feedback mechanisms
which couple the system's
dynamical performance (efficiency, sensitivity, robustness,...) with the
evolution of its structure. But the detailed interactions
between dynamics and evolution are often unclear
for real-world systems,  and understanding complex networks from the angle of
dynamics-structure interplay is still a challenging and
largely unexplored research topic.
Among the few theoretical works on dynamics-driven network evolutions
from the physics and the computer science communities 
(see, e.g.,
\cite{Skyrms-Pemantle-2000,Fabrikant-etal-2003,Barrat-etal-2004,Zimmermann-etal-2004,Schneider-Kirkpatrick-2005,Ehrhardt-etal-2006,Holme-Newman-2006,Oikonomou-Cluzel-2006,Garlaschelli-etal-2007,Xie-etal-2007,Kozma-Barrat-2008-b} and review \cite{Gross-Blasius-2008}), the main focus has been on network evolutionary games for which
network dynamics and evolution occur on comparable time scales. A payoff function is
defined for the system, and vertices
change their local connections to optimize gains. 
In many complex systems, however, the dynamical performance of a network
is a global property which can not be predicted by only looking at the local structures.
Most structural changes in such  systems, on the other hand, take place locally and
relatively randomly, without knowing their consequences to the system's dynamical performance.
The time-scales of network dynamics and network structural evolution can also be
very different. Will dynamics-structure coupling mechanisms build
highly nontrivial architectures out of random, blind, and local
structural mutations?

In this work extensive simulations of dynamics-driven network evolution
are performed on a simple model system, namely the local
 majority-rule (LMR) opinion dynamics of complex networks. 
There are two main motivations for this study. First,
earlier analytical and simulation studies
\cite{Zhou-Lipowsky-2005,Castellano-PastorSatorras-2006,Zhou-Lipowsky-2007}
revealed that networks with heterogeneous structural organizations have
remarkably better LMR dynamical performances than homogeneous networks.
In complementary to these studies, we want to know, in this
simple LMR dynamical system,
to what extent the dynamical performance of a network can influence
the evolution trajectory of the network's structure.
Second, as LMR-like dynamical processes are
frequently encountered in neural and gene-regulation networks and
other biological or social systems, it is hoped that a detailed study
of dynamics--structure coupling in the model LMR system will also
shed light on the structural evolution and optimization in real-world
complex systems. 

In the simulation, a fitness value is assigned
to a network to quantitatively measure its efficiency of LMR dynamical
performance. A slow (in comparison with the LMR dynamics)
mutation-selection process is performed on a population of networks,
and networks of higher fitness values are
more likely to remain in the population. The network
population dynamics reaches a steady state after
passing through several transient stages.
A steady-state network has
high clustering coefficient \cite{Watts-Strogatz-1998} and strong local degree-degree
correlations, and the fraction $P(k)$ of vertices in the network with degree $k$
resembles a power-law distribution of $P(k) \propto k^{-\gamma}$ with $\gamma \approx 2$.
Interestingly a global hub of degree proportional to network size $N$
spontaneously emerges in
the network. These results bring new insights on the
optimized network organization  for LMR dynamics. They are
also consistent with the opinion
that feedback mechanisms from dynamics to structure could be a dominant force
driving complex networks into  heterogeneous structures
\cite{Gross-Blasius-2008}.
Hopefully this work will stimulate  studies on the detailed interactions
between dynamics and structure in more realistic complex systems.

\section{Dynamics and evolution}

The local-majority-rule dynamics runs on a
network of $N$ vertices and $M = c N/2$ undirected links, with $c$ being the mean
connectivity. The network's adjacency matrix ${\bf A}$ has entries
$A_{ij} = 1$ if vertices $i$ and $j$ are connected by an edge or
$A_{ij} = 0$ if otherwise. Each vertex $i$ has an opinion (spin)
$\sigma_{i} = \pm 1$ that can be influenced by its nearest-neighbors.
At each time step $t$ of the LMR dynamics, every vertex of
the network updates its opinion synchronously according to
$\sigma_{i}(t + 1) = {\rm sign}[h_i(t)]$,
where 
$h_i(t) \equiv \sum_{j=1}^N A_{ij} \sigma_{j} (t)$
is the local field on vertex $i$ (when $h_i(t) = 0$ we set
$\sigma_i (t + 1) = \sigma_i (t)$). Starting from an initial configuration
${\vec{\sigma}}(0) \equiv  \{\sigma_1(0), \sigma_2(0),\ldots, \sigma_N(0) \}$,
the LMR process will derive the system to a consensus state in which all the vertices
share the same opinion. 
To measure  a network's efficiency of performing the LMR process, we follow
\cite{Zhou-Lipowsky-2005,Zhou-Lipowsky-2007} and choose the initial opinion patterns
${\vec{\sigma}}(0)$ to be {\em strongly disordered}, with
$\sum_{i=1}^N \sigma_i(0)= \sum_{i=1}^N k_i \sigma_i(0) \equiv 0$ ($k_i$ is the
degree of vertex $i$); in other words a
vertex (either randomly chosen or reached by following a
randomly chosen edge) has probability one-half to be in
the plus-opinion state. For networks containing $N\geq 1000$ vertices such
strongly disordered patterns can be easily constructed: One first divide the
vertices into two groups ($G_+$ and $G_-$) of equal size  
and assign spin $+1$ to vertices of group $G_+$ and spin $-1$ to
vertices of group $G_-$; and as long
as $\sum_{i=1}^{N} k_i \sigma_i \neq 0$,  two vertices (one from
$G_+$ and the other from $G_-$) are randomly chosen and their positions are
exchanged if and only if this exchange does not cause an increase in the
value of $| \sum_{i=1}^N k_i \sigma_i |$.  

For a given network ${\cal G}$, we generate a total number
$\Omega  = 1000$ of strongly disordered initial opinion patterns ${\vec{\sigma}}^\alpha(0)$
and, for each of them we run the LMR dynamics for one time step to reach the corresponding
pattern ${\vec{\sigma}}^\alpha(1)$. It has been shown in
\cite{Zhou-Lipowsky-2005} that the characteristic relaxation time of the LMR
dynamics is determined by the mean escaping velocity of the network's opinion pattern
from the strongly disordered region.
In the present work we calculate a fitness parameter $f$ for network ${\cal G}$ 
according to
\begin{equation}
	f({\cal G})= \frac{1}{\Omega} \sum\limits_{\alpha=1}^{\Omega}
\Biggl| \frac{1}{N} \sum\limits_{i=1}^{N} \sigma^\alpha_i(1) \Biggr| \ .
\label{eq:fitness}
\end{equation}
For networks with the same size $N$ and mean degree $c$,
we have checked that those with higher
fitness values have shorter mean LMR consensus times \cite{Shao-Zhou-2008-note}.

The fitness $f({\cal G})$ as defined by Eq.~(\ref{eq:fitness}) can also be
evaluated using completely random  configurations instead of
random strongly disordered configurations as the initial conditions.
When using random configurations
as the initial conditions, we found that the main results of the present
paper do not change, but the vertex-vertex correlation patterns of the
network will be slightly affected (i.e.,  the $R$ value defined by
Eq.~(\ref{eq:R}) will be more close to zero).

\begin{figure}
\includegraphics[width=0.6\textwidth]{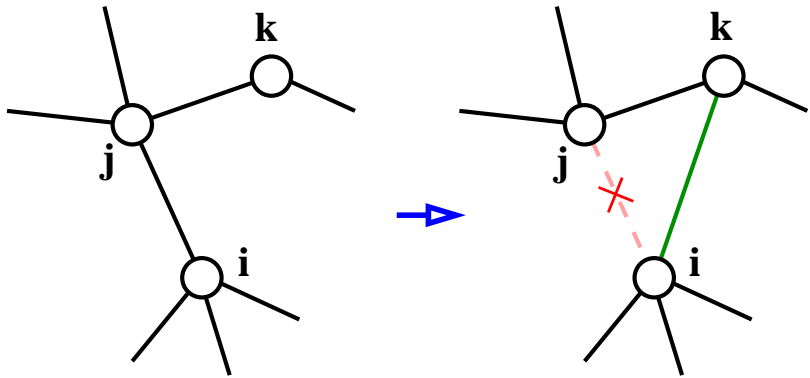}
\caption{\label{fig:edgemove}
Local edge redirection process. The edge rewiring $(i, j) \rightarrow (i, k)$
is accepted only if before the rewiring vertex $i$ and $k$ is not connected
and that vertex $j$ has vertex-degree higher than $k_0$.
}
\end{figure}

The network of a complex system is
not fixed but changes with time. We focus on
situations in which the typical time scale of network evolution
is much longer than that of the dynamical process.
In real-world systems, modifications of network structure
often occur distributedly and locally.
In accordance with this, in the simulation a simple local
edge-rewiring scheme as demonstrated in Fig.~\ref{fig:edgemove} is employed (similar simulation results obtained with
other local or global mutation schemes will be reported
elsewhere \cite{Shao-Zhou-2008-note}).
For each vertex $i$ of the network with probability
(mutation rate) $\mu$ propose the
following $(i,j)\rightarrow (i,k)$ edge redirection: randomly cut one of its edges $(i, j)$ to vertex $j$
and link this edge to vertex $j$'s nearest-neighbors
$k$. This proposal is rejected if (1) edge $(i, k)$ already exists
or (2) the degree of vertex $j$ is less than a minimal
value $k_0$ after cutting edge $(i, j)$. We set $k_0 = 3$ or $5$
in the simulation.  The only motivation of setting a cutoff $k_0>1$ is
to make sure
that in the network there will not be an accumulation of dangling vertices
of degree $k=1$. The local edge rewiring process reserves the number of
connected components of the network. If initially the network is
connected, it will remain to be connected. 

During each time step $T$ of the evolution,
the network is mutated by the above-mentioned random local
scheme and the fitness difference $\delta f$ between
the new and the old network is estimated by Eq.~(\ref{eq:fitness}).
For the dynamics-structure coupling, a simple simulation
rule could be to accept this mutation with probability unity
if $\delta f \geq 0$ and with probability $\exp(\beta \delta f)$
if otherwise, with $\beta$ controlling the strength of
dynamics-structure coupling. When there is only mutation but no
selection ($\beta = 0$), it has already been known that
the steady-state network's vertex-degree profile decays exponentially
for large degrees \cite{Baiesi-Manna-2003} (see also
the solid lines of Fig.~\ref{fig:DegreeDistribution}).
Here we focus on the other limit
of strong fitness selection $(\beta \gg 0)$ and carry out the network evolution
process through a population dynamics simulation of mutation and selection
\cite{Oikonomou-Cluzel-2006}: Starting with a set of $P = 25$ networks
uniformly sampled from the ensemble of random networks
of size $N$ and mean connectivity $c$, at each round of the
evolution each parent network generates $E = 3$ slightly
mutated offsprings, resulting in an expanded population
of $(E+1) P$ networks; the fitness values of these networks
are estimated and the $P$ ones with the highest values survive and
become parents for the next generation.
The population dynamics runs for many steps until the
system reaches a final steady-state. We have checked that
the steady-states of the simulation are not affect with larger
values of population parameters $P$ and $E$ \cite{Shao-Zhou-2008-note}.

\section{Results}

\begin{figure}
\begin{center}
\includegraphics[width=0.95\linewidth]{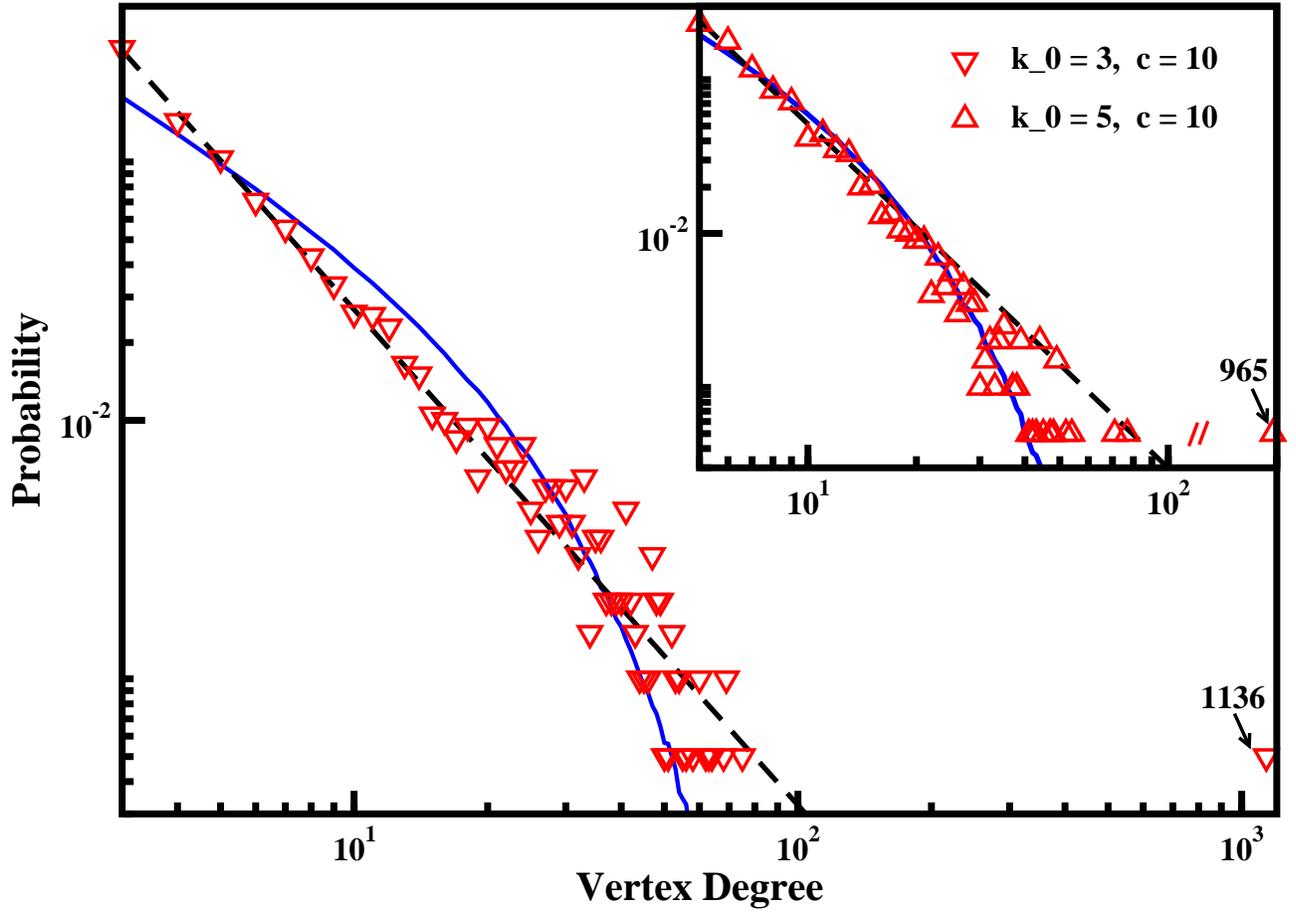}
\end{center}
\caption{(Color Online) The degree distribution for a steady-state network (under dynamics-structure coupling)
of $N=2000$ and $c=20$, $k_0=3$ (down triangles) and
$c=10$, $k_0=5$ (up triangles). The dashed lines are the best power-law fit of the points with
$\gamma=1.92\pm 0.02$ (main panel) and $\gamma=2.24 \pm 0.04$ (inset).
 The solid lines are the corresponding degree distributions (as averaged over
$200$ samples) for steady-steady networks under only mutation but no selection.
The mutation rate $\mu=0.01$.}
\label{fig:DegreeDistribution}
\end{figure}

Figure~\ref{fig:DegreeDistribution} shows the vertex-degree
distribution of a network with size $N=2000$ and mean degree $c$ ($=20$ or $10$)
at evolution time $T=2\times 10^5$ under mutation rate $\mu=0.01$.
One remarkable feature is that the steady-state network has a global hub whose connectivity is
proportional to $N$. This global hub 
samples the opinions of a finite population of the network (especially those of the
low-degree vertices, see below) and serves as a
global indicator of the system's state; its emergence is completely due to the dynamics-structure
coupling. To some extent this global hub balances
the influences of the minority high-degree vertices (see below) and those of the
majority low-degree vertices. Such a global hub may correspond to
news agencies and public medias in modern societies and to global transcription factors in
biological cells.

\begin{figure}
\begin{center}
\includegraphics[width=0.95\linewidth]{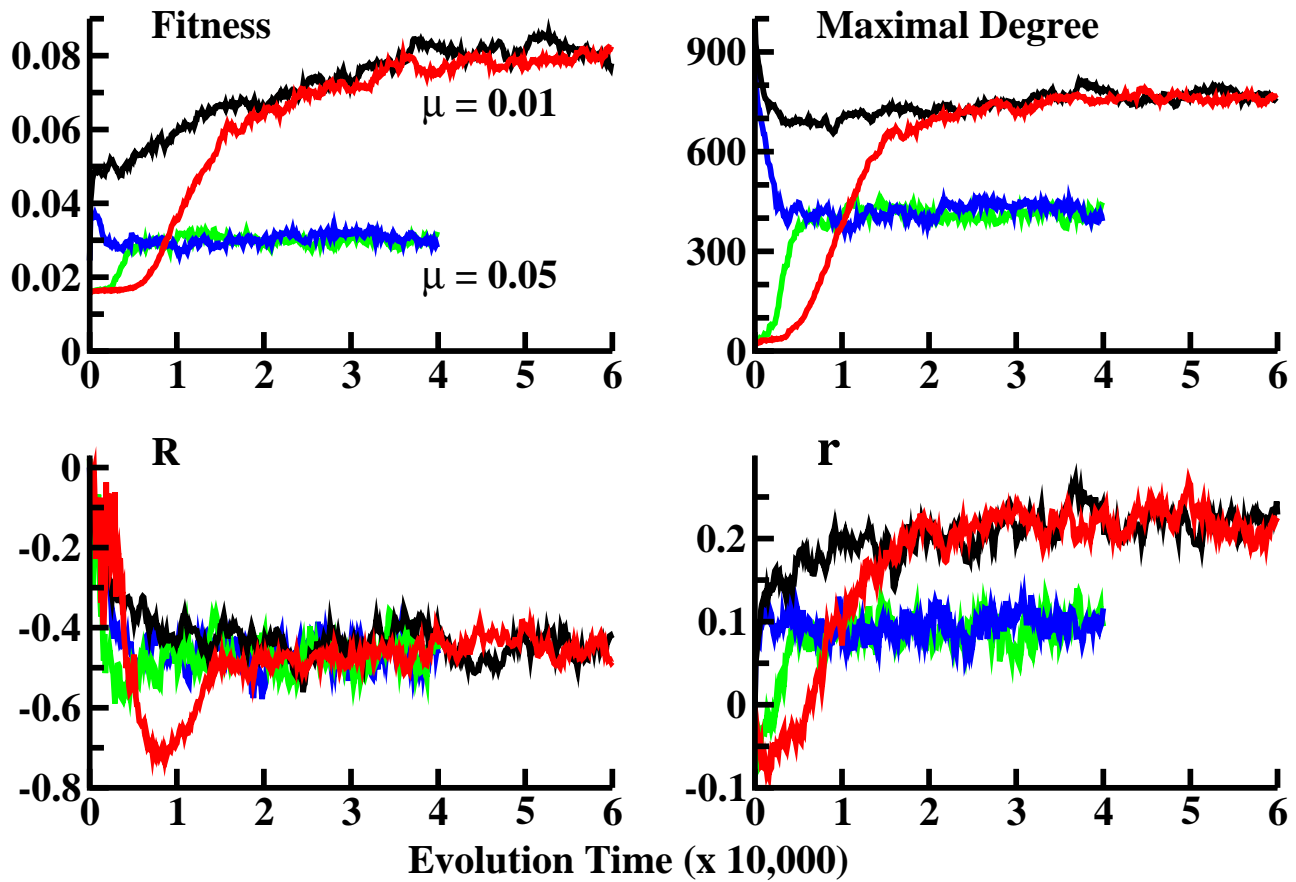}
\end{center}
\caption{(Color Online) The evolution of the mean fitness value,
the mean maximal vertex-degree, the correlation index $R$
[Eq.~(\ref{eq:R})], and the assortative-mixing index $r$ (of the
global hub-removed subnetwork) as a function of simulation steps.
The network size is $N = 1000$, mean vertex degree is
$c = 10.0$, and minimal degree is $k_0=5$. The mutation rate is $\mu = 0.01$ (black and red
curves, maximal evolution time to $6\times 10^4$ steps) or $\mu = 0.05$ (blue and green curves, maximal evolution time to $4 \times 10^4$ steps). The population dynamics
starts from an ensemble of random Poissonian networks
(red and green curves, which initially are the two lower curves) or an ensemble of
modified random Poissonian networks with a single vertex of
degree $N-1$ (black and blue curves, which initially are the two upper curves).}
\label{fig:Evolution}
\end{figure}

At mutation rate $\mu = 0.01$ the mean clustering coefficient \cite{Watts-Strogatz-1998}
of the steady-state networks is
about $0.0640\pm 0.0006$ (for $N=2000$ and $c=10$), considerably larger than the mean value of
$0.0048 \pm 0.0008$ for random Poissonian networks. The global hub is crucial for this
small-world property: if it is removed, the remaining subnetwork has much reduced
mean clustering coefficient $0.0089 \pm 0.0002$. There are strong degree-degree correlations
in a steady-state network. First, the
global hub prefers to interact with low-degree vertices. This preference can be measured by
a parameter $R$ defined by
\begin{equation}
R = \frac{ k^{{\rm g}}_{\rm nn} - \langle k^{{\rm g}}_{\rm nn} \rangle_{\rm rand} }
{\langle k^{{\rm g}}_{\rm nn} \rangle_{\rm rand} - k_{0} } \ ,
\label{eq:R}
\end{equation}
where $k^{{\rm g}}_{\rm nn}$ is the mean degree of nearest-neighbors of the
global hub and  $\langle k^{{\rm g}}_{\rm nn} \rangle_{\rm rand}$
is the averaged value of this mean degree over an ensemble of randomly shuffled networks.
The steady-state value is $R\approx -0.5$  for $N=1000$, $c=10$ and $\mu = 0.01$
(Fig.~\ref{fig:Evolution}). By this preference the $`$voices' of low-degree vertices
have a larger chance to be heard by the whole system.
Second, there are strong positive degree-degree correlations among vertices of a
steady-state network (excluding the global hub). To measure the extent of
these correlations, we calculate the assortative-mixing index $r$ of
the global hub-removed subnetwork following Ref.~\cite{Newman-2002}.
A steady-state assortative-mixing index of $r \approx 0.2$ for mutation
rate $\mu = 0.01$ (Fig.~\ref{fig:Evolution}) suggests that in the
subnetwork high-degree vertices (except the global hub) are more likely than random to
connect with other high-degree vertices.

The steady-state vertex degree distributions for networks under dynamics-structure
coupling deviate remarkably from those of the networks under only mutation
(see Fig.~\ref{fig:DegreeDistribution}).  Besides the emergence of a global
hub, the stead-state vertex degree distribution resembles a power-law form of
\begin{equation}
P(k) = C k^{-\gamma} \ , \hspace{0.5cm}k \geq k_0 \ ,
\label{eq:powerlaw}
\end{equation}
where $C$ is a normailzation constant.  For the data-set of Fig.~\ref{fig:DegreeDistribution} with
mean degree $c=20$ and minimal degree $k_0=3$, the fitting reports a 
decay exponent of $\gamma \simeq 1.92$, while for the data-set with of $c=10$ and
$k_0=5$ the fitting gives $\gamma \simeq 2.24$. 
Scale-free networks with decay exponent $\gamma < 2.5$ have shown to
be particularly efficient for LMR dynamics, and 
the mean relaxation time of the dynamics on such a network 
does not increase with network size $N$ \cite{Zhou-Lipowsky-2005,Zhou-Lipowsky-2007}.
This work indicates that such heterogeneous optimal network structures might be
reachable without employing any central planning and any intelligence.
The system only needs to
accumulate decentralized and local structural changes under the selection of dynamics-structure
coupling.
Networks with pronounced power-law degree distributions also emerged in
other model systems with comparable dynamicl and evolutionary time scales
\cite{Garlaschelli-etal-2007,Xie-etal-2007}.
In real-world systems, it was noticed by Aldana \cite{Aldana-2003} that
a major fraction of scale-free complex networks has their decay
exponent $\gamma$ in the tiny range of $\gamma \in [2.0, 2.5]$.

The network evolution trajectory also shows interesting
features. Starting from an ensemble of random Poissonian networks with
size $N=1000$ and mean degree $c=10.0$,
Fig.~\ref{fig:Evolution} shows that the evolution can be divided into
four stages. In the first $`$dormant' stage which lasts for about
$5000$ evolution steps for muatation rate $\mu = 0.01$, the degree distribution
of the networks changes gradually into the form shown by the solid line 
in the inset of Fig.~\ref{fig:DegreeDistribution}.
 The fitness values of the	
networks are  small and increase only very slowly, the
maximal vertex degrees of the networks are also small,
and the degree-degree correlations in the network are weak.
This dormant stage is followed by a $`$reforming' stage which lasts
for about $10,000$ steps for $\mu=0.01$. A global hub emerges and its degree
rapidly exceeds those of all the other vertices of the network, the subnetwork assortative
index $r$ also increases rapidly, and the degree distribution of the network becomes
power law-like at the end of this stage. This reforming stage has rapid increase in the
mean fitness value; it is follows by a long $`$structural fine-tunning' stage
(lasts for about $20,000$ steps at $\mu=0.01$) of slow increase in network fitness,
maximal degree, and assortative mixing. Finally
the network reaches the steady-state in which the
network's fitness value saturates but its local structures are being modified continuously.

We have performed simulations with different initial
conditions and confirmed that the steady-states are not affected
\cite{Shao-Zhou-2008-note}.
For example, Fig.~\ref{fig:Evolution} demonstrates that the steady-state networks
obtained from two different initial conditions share the same dynamical
performances and the same structural properties. (If the
network initially has a global hub of degree $N-1$ but otherwise is completely random,
during the evolution the degree of the global hub decreases but the fitness of the network
increases (Fig.~\ref{fig:Evolution}). This indicates that the
existance of a global hub, heterogeneous degree profile,
and strong local degree-degree correlations are all important for high dynamical
performance.) On the other hand, the evolution process is greatly influenced by the
network mutation rate $\mu$. For the same network size $N$ and
mean connectivity $c$, the steady-state networks obtained with a lower network mutation rate
$\mu$ have better dynamical performances (Fig.~\ref{fig:Evolution}). 
As the network topology becomes heterogeneous, most local structural changes will tend to
deteriorate the dynamical performance. When the mutation
rate is relatively large, in one evolution step the probability for
the combination of $L= N \mu$ local and distributed mutations to enhance
the network's dynamical performance will decrease rapidly with $L$. 
The balance between structural entropy (randomness)
and dynamical performance then makes the system cease to be further optimized. For the
dynamics-structure interaction to work most efficiently, it is
therefore desirable that the time scale of network evolution be much slower than the
time scale of network dynamics.

\section{Conclusion}

In this paper, we have studied the evolution and
optimization of complex networks from the
perspective of dynamics-structure mutual influence. Through
extensive simulation on a simple prototypical model process,
the local-majority-rule dynamics,
we showed that if there exist feedback mechanisms from a network's
dynamical performance to its structure, the network can be driven into 
 highly heterogeneous structures with a global hub,
strong local correlations in its connection pattern, and  power law-like
vertex-degree distributions. The steady-state networks
reached by this dynamics-driven evolution will have better dynamical
performance if network evolution occurs much slowly than the dynamical process on the
network.

For the LMR dynamics specifically, this work confirmed and extended previous studies
\cite{Zhou-Lipowsky-2005,Zhou-Lipowsky-2007} by showing that scale-free networks with
decay exponent $\gamma < 2.5$ indeed are
optimal and can be reached without the need of any central planning. Besides the
scale-free property and strong local structural correlations, a steady-state network
also has a global hub which serves a global indicator of the system's state by
sampling the opinions of a large fraction of the
vertices of the system. 

Real-world complex systems of cause are much more complicated than the simple
model systems studied in this paper. Different mechanisms may be contributing
simultaneously to the evolution of a real-world complex network.
The present paper suggested that
the interplay between dynamics and structure can be an
 important driving force for the formation and stabilization
of heterogeneous structures which are
ubiquitous in biological and social systems
\cite{Skyrms-Pemantle-2000,Barrat-etal-2004,Zimmermann-etal-2004,Ehrhardt-etal-2006,Holme-Newman-2006,Garlaschelli-etal-2007,Xie-etal-2007,Kozma-Barrat-2008-b,Gross-Blasius-2008}.
A lot of efforts are needed to decipher the detailed dynamics-structure coupling
mechanisms in many complex systems.

\section*{Acknowledgement}

We thank Kang Li and Jie Zhou for helpful discussions, 
Erik Aurell and Peter Holme for their suggestions on the manuscript, and Zhong-Can Ou-Yang
for support.
We benefited from the KITPC 2008 program $``$Collective
Dynamics in Information Systems''. The State Key Laboratory for Scientific and 
Engineering Computing, CAS, Beijing is kindly acknowledged for computational facilities.

\end{document}